\documentclass{aa}

\usepackage{graphicx}
\usepackage{natbib}
\bibpunct{(}{)}{;}{a}{}{,}
\begin{document}
\title{On a probable manifestation of Hubble expansion \\
       at the local scales, as inferred from LLR data}

\titlerunning{Hubble expansion at local scales, as inferred from LLR data}

\author{Yu.V. Dumin}

\offprints{Yu.V. Dumin}

\institute{IZMIRAN, Russian Academy of Sciences,
           Troitsk, Moscow reg., 142190 Russia \\
           \email{dumin@yahoo.com, dumin@cityline.ru}
}

\date{Received March 11, 2002; accepted Month Day, 2002}

\abstract{
Processing the data of lunar laser ranging (LLR), whose accuracy
now reaches a few millimeters, points to the effect of anomalous
increase in the lunar semimajor axis (with an excessive rate
1.3 cm/yr), which cannot be attributed to the well-known tidal
exchange of angular momentum between the Earth and Moon.
One of the possible interpretations of the above-mentioned anomaly may be
the ``local'' Hubble expansion with
$ H_0^{\mathrm{(loc)}} = 33 \pm 5 $\,(km/s)/Mpc.
This small value of the local Hubble constant (about two times less than
at intergalactic scales) can be reasonably explained if it is formed only
by some kind of an unclumped ``dark matter'' or ``dark energy'',
while the other kinds of matter experienced a gravitational instability,
formed compact objects, and no longer contribute to the formation of
Hubble expansion at the local scales.

\keywords{
Relativity --
cosmological parameters --
dark matter --
celestial mechanics --
Earth --
Moon }
}

\maketitle

\section{Introduction}

Precise measurements of the Earth--Moon distance by using
the ultrashort laser pulses -- the lunar laser ranging (LLR) -- are
carried out for over 30 years, after the installation of
several retroreflectors on the lunar surface in the course of
Apollo (USA) and Lunakhod (USSR, in collaboration with France)
space missions \citep[e.g. reviews by][]{Dic94,Nor99,Sam98}.
The typical accuracy of these measurements was:
\begin{itemize}
\item ${\sim}25$\,cm  in the early 1970's,
\item 2--3\,cm  in the late 1980's,  and
\item a few millimeters  at the present time.
\end{itemize}

LLR works contributed significantly to astrometry, geodesy, geophysics,
lunar planetology, and gravitational physics.
The most important results related to General Relativity are:
\begin{itemize}
\item verification of the Strong Equivalence Principle
with accuracy up to ${\sim}10^{-13}$,
\item determination of the first post-Newtonian parameters in
the gravitational field equations,
\item detection of the so-called geodetic precession of the lunar orbit, and
\item imposing the observational constraints on time variations
in the gravitational constant $ {\dot{G}} / G $ with accuracy
${\sim}10^{-11}$ per year.
\end{itemize}

\section{Using the LLR data for finding the ``local'' Hubble constant}

Despite the considerable advances listed above,
there is a long-standing unresolved problem in the interpretation of
LLR data -- anomalous increase in the lunar semimajor axis
\citep[e.g.][]{Per00}.

\begin{figure}
\centering
\includegraphics{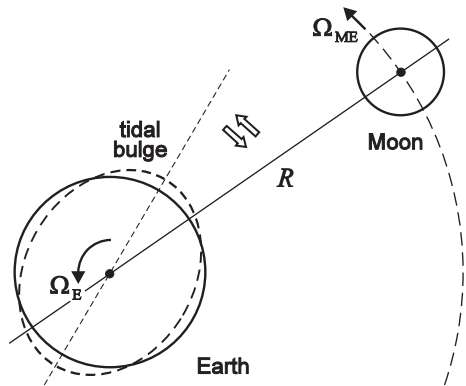}
\caption{A scheme of tidal exchange of angular momentum
         between the Earth and Moon.}
\label{Tidal_Inter}
\end{figure}

In general, such increase is well-known and can be
\emph{partially} explained by tidal interaction
between the Earth and Moon \citep[e.g.][]{Kau68}.
Because of the relaxation processes,
the tidal bulge is not perfectly symmetric about the Earth--Moon line
but slightly shifted in the direction of Earth's rotation
(see Fig.~\ref{Tidal_Inter}).
As a result, there is a torque moment,
which decelerates a proper rotation of the Earth
and accelerates an orbital rotation of the Moon;
so that the mean Earth--Moon distance increases.

From the angular momentum conservation law
\begin{equation}
I_{\mathrm{\scriptscriptstyle{E}}}
\frac{\mathrm{d}}{\mathrm{d}t} {\Omega}_{\mathrm{\scriptscriptstyle{E}}} +
m_{\mathrm{\scriptscriptstyle{M}}} \frac{\mathrm{d}}{\mathrm{d}t}
\left( R^2 {\Omega}_{\mathrm{\scriptscriptstyle{ME}}} \right) = 0
\end{equation}
and the relation between the lunar orbital velocity and
its distance from the Earth
\begin{equation}
{\Omega}_{\mathrm{\scriptscriptstyle{ME}}} =
G^{1/2} \, m_{\mathrm{\scriptscriptstyle{E}}}^{1/2} \, R^{-3/2} ,
\end{equation}
it can be easily found that
the rate of increase in the lunar semimajor axis $\dot{R}$
is related to the rate of deceleration of the Earth's rotation
${\dot{T}}_{\mathrm{\scriptscriptstyle{E}}}$ by the well-known formula:
\begin{eqnarray}
\dot{R} & = & k \, {\dot{T}}_{\mathrm{\scriptscriptstyle{E}}} \, ,
\label{R-T}
\\
k & = & 4 \, \pi \, G^{-1/2} \, I_{\mathrm{\scriptscriptstyle{E}}} \,
m_{\mathrm{\scriptscriptstyle{E}}}^{-1/2} \,
m_{\mathrm{\scriptscriptstyle{M}}}^{-1} \, R^{1/2} \,
T_{\mathrm{\scriptscriptstyle{E}}}^{-2}
\nonumber \\
& = & 1.81{\cdot}10^5 \, \mathrm{cm/s} ,
\end{eqnarray}
where
${\Omega}_{\mathrm{\scriptscriptstyle{E}}}$ and
$T_{\mathrm{\scriptscriptstyle{E}}}$ are the angular velocity and period
of the proper rotation of the Earth,
${\Omega}_{\mathrm{\scriptscriptstyle{ME}}}$ is the angular velocity
of orbital rotation of the Moon about the Earth,
$R$ is the distance between them%
\footnote{Within the accuracy required here,
we can neglect the ellipticity of the lunar orbit.},
$I_{\mathrm{\scriptscriptstyle{E}}}$ is the Earth's moment of inertia,
$m_{\mathrm{\scriptscriptstyle{E}}}$ and
$m_{\mathrm{\scriptscriptstyle{M}}}$ are the terrestrial and lunar masses,
and $G$ is the gravitational constant.

As follows from \emph{telescopic observations} of the Earth's rotation
with respect to the distant objects, collected over the last 300~years,
$\dot{T}_{\mathrm{\scriptscriptstyle{E}}}^{\mathrm{(tel)}} \! = \!
1.4{\cdot}10^{-5}${\,}s/yr \citep{Per00}.
So, according to Eq.~(\ref{R-T}),
$\dot{R}^{\mathrm{(tel)}} = 2.53${\,\,}cm/yr.
On the other hand, immediate measurements of the Earth--Moon distance 
by \emph{LLR technique} give an appreciably greater value
$\dot{R}^{\mathrm{(LLR)}} = 3.82${\,\,}cm/yr \citep{Dic94}.

The most of attempts to explain the anomaly
$\Delta \dot{R} = \, \dot{R}^{\mathrm{(LLR)}} -
\, \dot{R}^{\mathrm{(tel)}} = 1.29${\,\,}cm/yr
were based on accounting for some additional geophysical effects
(e.g. secular changes in the Earth's moment of inertia
$I_{\mathrm{\scriptscriptstyle{E}}}$).
Unfortunately, they did not lead to a satisfactory quantitative agreement
with observations.

Yet another promising interpretation of the above-mentioned
anomaly $\Delta \dot{R}$, from our point of view,
is Hubble expansion in the local space environment,
which should contribute to $\dot{R}^{\mathrm{(LLR)}}$
but will not manifest itself in $\dot{R}^{\mathrm{(tel)}}$.
As follows from the standard relation
$\Delta \dot{R} = H_0^{\mathrm{(loc)}} R$,
the ``local'' Hubble constant should be
\begin{equation}
H_0^{\mathrm{(loc)}} = \, 33 \pm 5 \, \mathrm{(km/s)/Mpc}.
\label{H_loc}
\end{equation}

\section{Discussion}

At first sight, the Hubble constant given by Eq.~(\ref{H_loc})
seems to be erroneous, since it equals only about one-half
the commonly-accepted value at intergalactic scales.
Nevertheless, it can be reasonably interpreted if
the local Hubble expansion is formed only by some kind of
an \emph{unclumped} ``dark matter'' or ``dark energy'',
uniformly distributed in the Universe
(such as $\Lambda$-term, ``quintessence'', inflaton-like scalar field,
and so on);
while the other kinds of matter experienced a gravitational instability,
formed compact objects, and are no longer able to make their contributions
to the rate of Hubble expansion at the local scales.

Besides, apart from any theoretical arguments,
the recent observations \citep[e.g. by][]{Ekh01} revealed that
a linear ``quiescent'' Hubble flow begins at least from the distances
${\sim}$1--2{\,}Mpc, i.e. an order of magnitude less
than was usually expected before.
This fact was also interpreted by \citet{Che01} as
a manifestation of some unclumped dark matter.

So, the LLR technique, which was used so far only in solar-system studies,
may also become a valuable tool for solving the cosmological problems,
because it enables us either to measure a local rate of Hubble expansion or
to impose an upper limit on this quantity \citep{Dum01gra}.

Of course, a cosmological nature of the anomalous increase
in the Earth--Moon distance is not reliably established by now.
It may be also caused merely by some geophysical artefacts.
To distinguish between these two possibilities,
it would be very interesting to seek a similar effect
in some artificial satellite system, which does not suffer from
geophysical uncertainties \citep{Dum01lanl}.
The suitable objects may be the space-based laser interferometers
projected for searching the gravitational waves,
such as LISA \citep[e.g.][]{Ben98}.

\begin{acknowledgements}
I am very grateful to
Yu.V.~Baryshev,
P.L.~Bender,
A.D.~Chernin,
E.V.~Derishev,
C.~Hogan,
S.M.~Ko\-peikin,
S.M.~Molodensky,
M.~Tinto, and
A.V.~To\-po\-rensky
for valuable discussions and critical comments,
as well as to
D.P.~Kirilova and
V.D.~Kuznetsov
for various help in the course of work.
\end{acknowledgements}

\bibliographystyle{aa}

\end{document}